\documentstyle[preprint,12pt,aps]{revtex}

\begin{document}
\draft
\title{Defect Energy with Conjugate Boundary Conditions in Spin Glass Models 
       in Two Dimensions}

\author{F. Matsubara}
\address{Department of Applied Physics, Tohoku University, Sendai 980-8579,
Japan}

\author{T. Shirakura}
\address{Faculty of Humanities and Social Sciences, Iwate University, 
Morioka 020, Japan}

\author{M. Shiomi}
\address{Department of Applied Physics, Tohoku University, Sendai 980-8579,
Japan}

\date{\today}
%
\maketitle
\begin{abstract}

We calculate the naive defect energy $\Delta E$ of Ising spin glass(SG) 
models in two dimensions using conjugate boundary conditions.  
We predict that, in the $\pm J$ model, the averaged value 
$\overline{\Delta E}$ converges to some non-zero value in the thermodynamic 
limit in contrast with $\overline{\Delta E} = 0$ in the Gaussian model. 
This prediction is incompatible with previous ones but supports a recent 
Monte Carlo prediction of the presence of the SG phase at finite temperatures 
in the $\pm J$ Ising model. 
We also calculate the interface free energy to confirm it.

\end{abstract}



Spin glasses have attracted great challenge for computational physics in 
these two decades. It is widely believed that in two dimensions 
the spin glass(SG) transition occurs at zero temperature $T_c = 0$. 
This belief arises from the study of the stiffness exponent $\theta_S$ 
which has a positive or negative value 
when the phase transition occurs at a finite, non-zero temperature 
$T_c \neq 0$ or $T_c = 0$, respectively. 
McMillan\cite{McMillan}, and Bray and Moore\cite{BrayM} estimated 
a value of $\theta_S \sim -0.28$ for the Ising model with Gaussian 
distribution of bonds(Gaussian model) by calculating the defect energy 
$\Delta E$ of finite lattices and predicted that 
the model exhibits the SG transition at $T_c = 0$ with the correlation 
exponent of $\nu = -1/\theta_S \sim 3.5$.  
This value of $\theta_S$ was confirmed by a recent estimation using larger 
lattices\cite{Rieger}. 
It should be noted, however, that the value of $\nu$ is significantly 
different from direct estimations of 
$\nu \sim 2.0$\cite{KawashimaHS,Liang,NeyY}. 
The problem is subtle in a discrete model with $+J$ and $-J$ bonds($\pm J$ 
model) for which the value of $\theta_S$ was firstly estimated 
as $\theta_S \sim 0$\cite{Ozeki1,Cieplak,Ozeki2}. 
However, a recent estimation gave a small negative value of 
$\theta_S \sim -0.05$\cite{KawashimaR}. 
From these results and an idea that the nature of the phase transition at a 
finite temperature would not depend on details of the bond distribution, 
one believes that {\it $T_c = 0$ in any two dimensional SG model}, 
although no direct evidence of $T_c = 0$ has yet been obtained for every 
model, especially for the $\pm J$ model\cite{Comm1}. 
Recently, Shirakura and Matsubara made a Monte Carlo simulation of the $\pm J$ 
model at low temperatures\cite{SM1} and that of an asymmetric discrete model 
with $+J$ and $-aJ(a \neq 1)$ bonds at very low temperatures\cite{SM2} 
and predicted $T_c \neq 0$. 
Obviously, their prediction and the belief are incompatible.


In this Letter, we carefully reexamine the defect energy of Ising SG models 
in two dimensions and find that the results support the prediction of 
$T_c \neq 0$. 
Our findings are as follows. 
(i) The conventional boundary conditions used for estimating the defect 
energy $\Delta E$  are inadequate in some cases, especially in delicate 
problems such as the phase transition of the $\pm J$ model. 
{\it A naive defect energy is calculated for the first time in the SG problem 
using conjugate boundary conditions.} 
(ii) The distribution of $\Delta E$, $P(\Delta E)$, is highly asymmetric 
for small systems and {\it the conventional estimation of $\theta_S$ by means 
of the size dependence of the average value $\overline{\Delta E}$ could take 
the risk for misleading the conclusion.} 
(iii) $\overline{\Delta E}$'s are estimated for both Gaussian model and 
discrete models for different sizes of the lattice and the following 
predictions are given for large systems. 
In the Gaussian model, $P(\Delta E)$ shrinks to a sharp one at $\Delta E = 0$.
Therefore $\overline{\Delta E} = 0$ and $\theta_S$ is negative as predicted 
previously but its value is considerably larger than that of the previous 
estimations. 
In the discrete models, $P(\Delta E)$ exhibits discrete peaks whose 
weights at $\Delta E \sim 0$ as well as those at large $\Delta E$ never 
increase with increasing the size of the lattice.  
Therefore $\overline{\Delta E}$ converges to some non-zero value, i.e., 
$\theta_S = 0$, in contrast with the recent prediction. 
(iv) Thus {\it the prediction of $T_c \neq 0$ for the discrete models is not 
incompatible with the results of the defect energy analysis.} 
The interface free energy is also calculated to confirm the presence of 
the SG phase at finite temperatures.


We start with an Ising model on a square lattice $L \times (L+1)$ described by 
the Hamiltonian 
\begin{eqnarray} 
     H = - \sum_{<i,j>}J_{ij}\sigma_{i}\sigma_{j}, 
\end{eqnarray} 
where $\sigma_{i} (= \pm 1)$ are Ising spins and $<ij>$ runs all nearest 
neighbor pairs. We consider the following two bond distributions:
\begin{eqnarray}
    P(J_{ij}) &=& \frac{1}{\sqrt{2\pi}}\exp{(-J_{ij}^2/2)}  \\
    P(J_{ij}) &=& \frac{1}{2}[\delta(J_{ij} - J) + \delta(J_{ij} + aJ)]. 
\end{eqnarray}
The model(1) with the distribution (2) is the Gaussian model and that 
with (3) is a discrete model. Hereafter we call the direction 
for $(L+1)$ spins as the $x$-direction and the other as the $y$-direction. 
The defect energy $\Delta E$ has been conventionally defined by 
$\Delta E = E_{ap}-E_p$, where $E_{p}$ and $E_{ap}$ are the ground state 
energies for the periodic and anti-periodic boundary conditions 
in the $y$-direction. 
The boundary condition in the $x$-direction is chosen to be either 
periodic\cite{McMillan} or free\cite{Rieger,KawashimaR}. 
Note that the defect energy calculated in this way is either positive or 
negative and one consider the absolute value $|\Delta E|$. 
Bray and Moore\cite{BrayM} applied somewhat different boundary conditions. 
They considered the lattice with the periodic boundary condition in the 
$y$-direction. 
The spin configurations on the surfaces in the $x$-direction, which are 
denoted as \{$S_a$\} for $x = 1$ and \{$S_b$\} for $x = (L+1)$, are put 
at random. 
They defined $E_p$ as the ground state energy of the lattice with \{$S_a$\} 
and \{$S_b$\}, and $E_{ap}$ as that with  \{$S_a$\} and \{$-S_b$\}, where 
\{$-S_b$\} is the spin configuration obtained reversing all the spins 
of \{$S_b$\}. 
These different sets of boundary conditions would be essentially the same 
for evaluating $\Delta E$. 
In fact, the same value of $\theta_S \sim -0.28$ has been obtained in the 
Gaussian model\cite{McMillan,BrayM,Rieger}. 
However, it is not obvious whether these sets give the true defect 
energy or not. 
For example, we consider the boundary conditions by Bray and Moore. 
For neither boundary condition, the system will not have its ground state spin 
configuration of the lattice without any restriction. 
That is, some defect lines (or defect points) already exist in those ground 
states. 
Therefore, it is doubtful whether $\Delta E$ gives true defect energy or not. 
This problem would not be so serious when $\Delta E$ has a strong dependence 
on $L$ like in the ferromagnetic case. 
However, in the two dimensional SG model, the size dependence is slight, e.g., 
$\Delta E \sim J$ even for $L \sim 30$\cite{Rieger,KawashimaR}. 
That is, the value would considerably change even when only one position of 
the defect line changes. 
To relieve this difficulty, Ozeki\cite{Ozeki2} used a replica boundary 
condition for one end  but still used the fixed boundary condition for 
the other end. 
It should be noted that he obtained a value of $\theta_S$ slightly larger 
than that of the conventional method\cite{Ozeki1,Ozeki2}.


We consider the lattice treated by Bray and Moore. 
It is obvious that the spin configuration without any defect is the ground 
state without any restriction. 
Then it is quite natural to choose the spin configurations \{$S_a$\} and 
\{$S_b$\} as ones in the ground state for the free boundaries. 
We call this boundary condition a conjugate boundary condition, because it 
gives the true ground state energy. 
By the use of the boundary conditions of \{$S_a$\} and \{$-S_b$\}, 
we may certainly construct one defect line, if it could occur. 
The defect energy $\Delta E$ obtained using these sets of boundary conditions 
is, of course, non-negative in contrast to the conventional one. 
The problem is how to get \{$S_a$\} and \{$S_b$\}. 
We can readily get them by using a cluster heat bath(CHB) 
method\cite{SM2,CHB1,CHB2} for both $T = 0$ and $T \neq 0$. 
Some comments should be given. 
For the Gaussian model, the ground state can be uniquely determined. 
Then $\Delta E$ for each sample can be uniquely determined. 
On the other hand, in the discrete models, there are many different sets 
of \{$S_a$\} and \{$S_b$\}. 
Then we choose one of them for \{$S_a$\}, and \{$S_b$\} is chosen so as to 
give the minimum value of $\Delta E$\cite{Comm2}. 
These calculations may be readily done by using the transfer matrix 
method\cite{Ozeki1,MorgenB}.


We make the simulation using these conjugate boundary conditions. 
The lattices treated here are $L \times (L+1)$ for $L \leq 24$ and the 
numbers of the samples are $N_s = 10,000 \sim 50,000$. 
For every size of the lattice with $L$, the defect energy $\Delta E$ is 
calculated for every sample and the distribution function $P_L(\Delta E)$ 
is obtained. 
The defect energy $W(L)( \equiv \overline{\Delta E})$ for the lattice with 
$L$ is obtained from $W(L)  = \int \Delta E P_L(\Delta E) d\Delta E$.

The results of $W(L)$ are shown in Fig. 1 in a log-log scale. 
These size dependences of $W(L)$ are quite different. 
In the Gaussian model, $\log W(L)$ decreases almost linearly with $\log L$ 
suggesting $W(L) \sim L^{\theta_S}$ with $\theta_S < 0$ 
as predicted previously. 
However, the value of $\theta_S \sim -0.20$ is considerably larger 
than that of the previous estimations of $\theta_S \sim 
-0.28$\cite{McMillan,BrayM,Rieger}. 
In the $\pm J$ model, $W(L)$ slightly increases in contrast with the 
recent prediction\cite{KawashimaR}. 
On the other hand, in the asymmetric model with $+J$ and $-0.8J$ bonds, 
$W(L)$ decreases.  
These results in the descrete models are quite mysterious, because the MC 
studies\cite{SM1,SM2} suggested $T_c \neq 0$ for both the models.

To examine this problem, we consider $P_L(\Delta E)$ itself. 
In Fig. 2, we present $P_L(\Delta E)$ of the Gaussian model. $P_L(\Delta E)$ 
has a continuous weight in a finite range of $\Delta E$. 
As $L$ is increased, the weight at $\Delta E \sim 0$ increases, while that 
at larger $\Delta E$ decreases.  
This fact suggests that it collapses to $P_L(\Delta E) \sim \delta(\Delta E)$. 
On the other hand, in the $\pm J$ model, $\Delta E$ takes values for every 
$4J$, i.e., $P_L(\Delta E) = \sum_{l=0,4,\cdots} A_l \delta(\Delta E - lJ)$. 
The size dependencies of the coefficients of $A_l$ are plotted in Fig. 3. 
Only $A_0$ and $A_4$ have considerable weights. 
The most important point is that $A_0$ and $A_{4n}$ with $n \geq 2$ never 
increase with $L$. 
This means that $W(L)$ never vanishes nor diverges for $L \rightarrow \infty$, 
i.e., $W(L)$ converges to some non-zero value. 
In the asymmetric model, $P_L(\Delta E)$ also has discrete peaks at 
every $2(1-a)J$. 
In Fig. 4, we show $P_L(\Delta E)$ in a line graph. 
For small $L$, it has a doubl peak at $\Delta E \sim 0.4J$ and $3J$. 
As $L$ is increased, the peak at $\Delta E \sim 3J$ rapidly diminishes and 
$P_L(\Delta E)$ becomes of the single peak. 
The weights of $\Delta E = 0$ and $0.4J$ components decrease and those at 
$\Delta E \sim J$ increase. 
Thus it is suggested that, for $L \rightarrow \infty$, $P_L(\Delta E)$ has 
a single peak at $\Delta E \sim J$. 
That is, $W(L)$ also converges to some non-zero value. 
The distribution functions $P(\Delta E)$ for $L \rightarrow \infty$ suggested 
above are schematically shown in Fig. 5.

We conclude, hence, that whether the SG phase transition occurs or not 
in two dimensions depends on the model. 
In the Gaussian model, $T_c = 0$ as predicted previously. 
It should be emphasized again that the value of $\theta_S$ is considerably 
different from that of the previous estimation. 
The value seems to have no relation with that of the correlation exponent 
$\nu$\cite{KawashimaHS}. In the discrete models, the problem is very delicate, 
because $W(\infty)$ has a finite, non-zero value 
like that in the two dimensional $xy$ ferromagnet. 
This result suggests that $T = 0$ is marginally stable. 
The SG order will exist at $T = 0$ which is characterized by a power 
law decay of the spin correlation\cite{Ozeki1,MorgenB}. 
Whether $T_c \neq 0$ or not should be examined separately. 
We have also made it by evaluating the interface free energy 
$\Delta F = F_{ap} - F_p$\cite{McMillan,Ozeki2}, where $F_p$ and $F_{ap}$ 
are the free energies calculated using conjugate boundary conditions 
\{$S_a$\} and \{$S_b$\} at finite temperatures\cite{Comm3}. 
The result for the $\pm J$ model is presented in Fig. 6. 
In fact, the average value $\overline{\Delta F}$ for $T \leq 0.1J$ seems to 
increase with lattice size $L$, while that for $T \geq 0.3J$ decreases. 
Thus we believe that the phase at $T = 0$ persists up to some finite 
temperature. 
That is, $T_c$ is non-zero and exists between $0.1J$ and $0.3J$, 
probably $T_c \sim 0.2J$. 
This result is quite interesting, because the value of $T_c \sim 0.2J$ is 
compatible with the previous estimation of $T_c \sim 0.24J$ using the MC 
method\cite{SM1}. 
Thus we believe that the defect energy analysis gives results 
which are not incompatible with the MC result.

Finally, we should note that the choice of the boundary condition will be 
crucially important also in three dimensions. 
Especially, studies of the defect energy in the three dimensional vector 
SG models under the conjugate boundary conditions are desirable, 
because the possibility of the chiral SG ordering without any spin ordering 
is a current topic\cite{Kawamura1,Kawamura2,MaucourtG}.

\bigskip

One of the authors (TS) wish to thank Professor H. Takayama, Dr. K. Hukushima, 
Dr. H. Yoshino and Dr. S. Todo for valuable discussions. 
The simulations were made on SX4 at the Computer Center of Tohoku University.




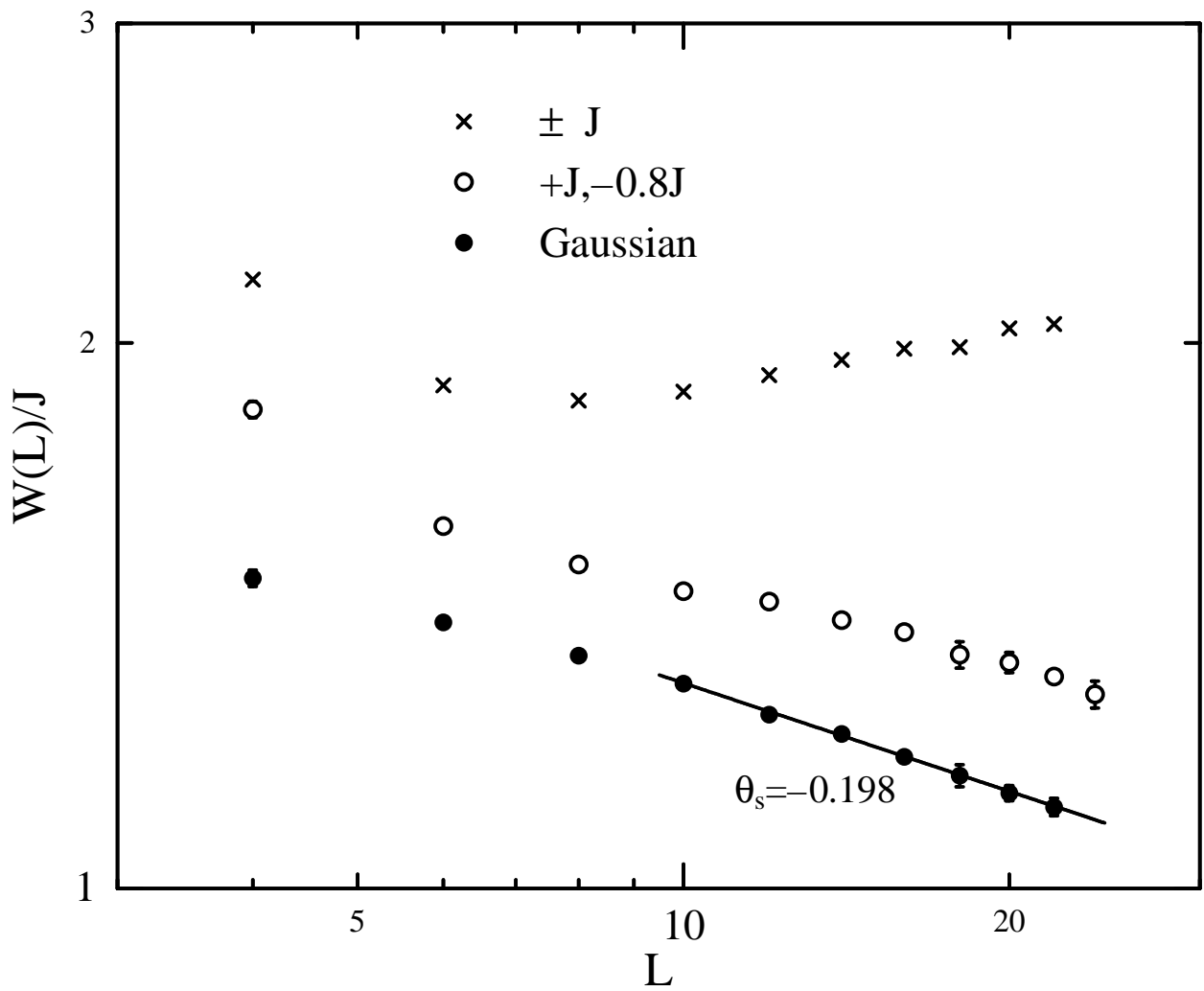
\begin{figure}
\caption{The defect energies $W(L)$ as functions of the lattice size $L$.}
\end{figure}

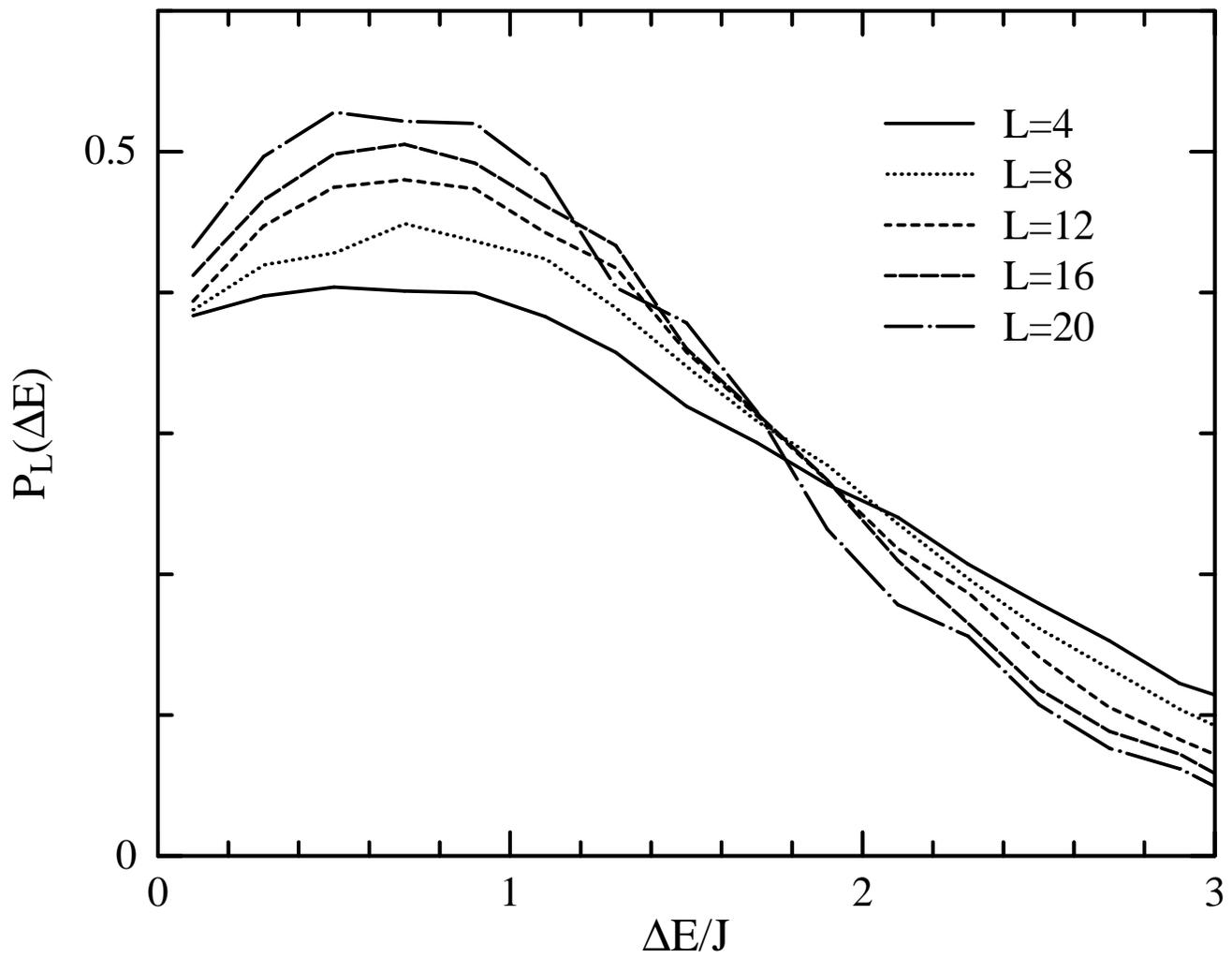
\begin{figure}
\caption{Distribution of $\Delta E$ of the Gaussian model.}
\end{figure}

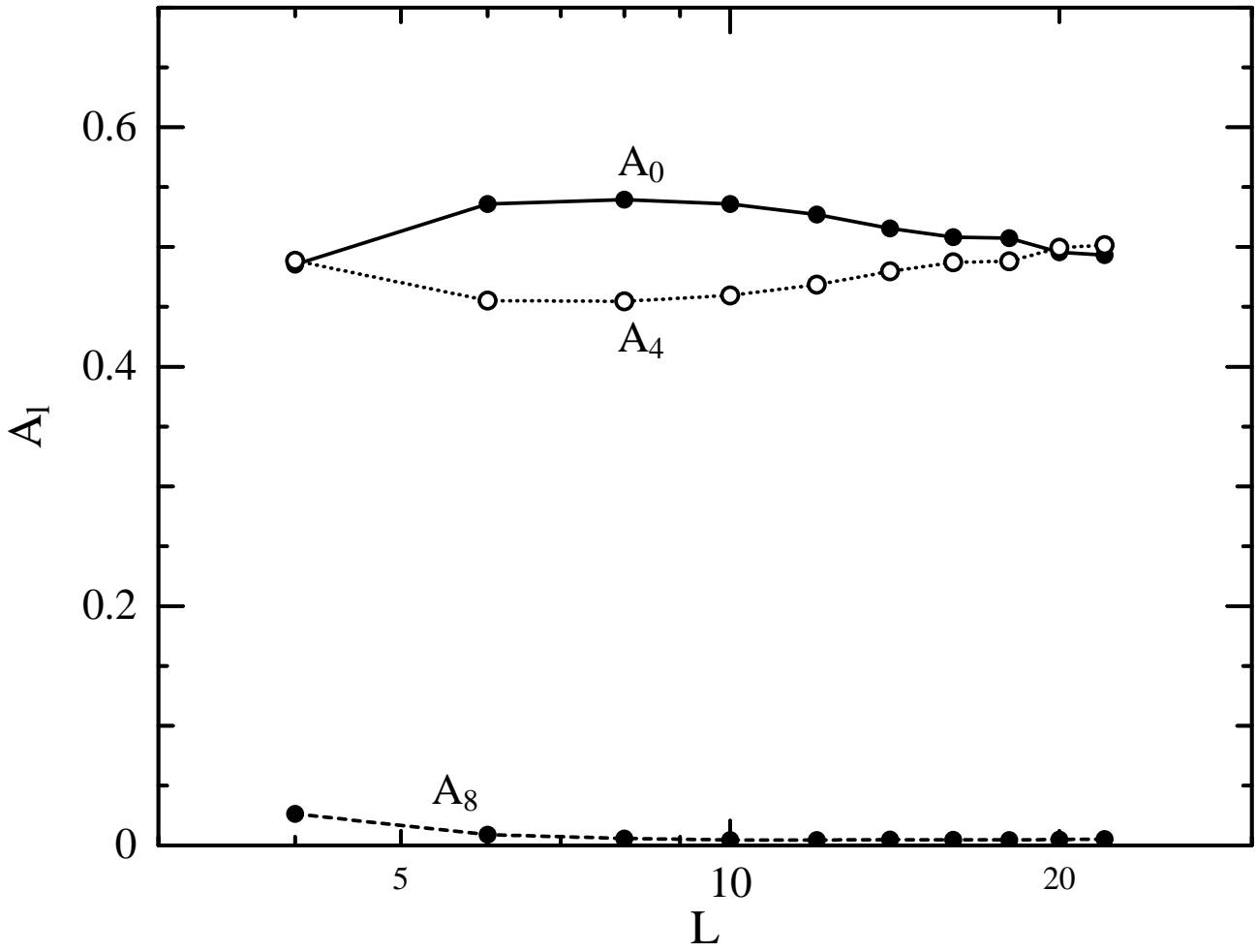
\begin{figure}
\caption{The weights $A_l$ of discrete peaks at $\Delta E = lJ$ of the $\pm J$ 
         model.}
\end{figure}

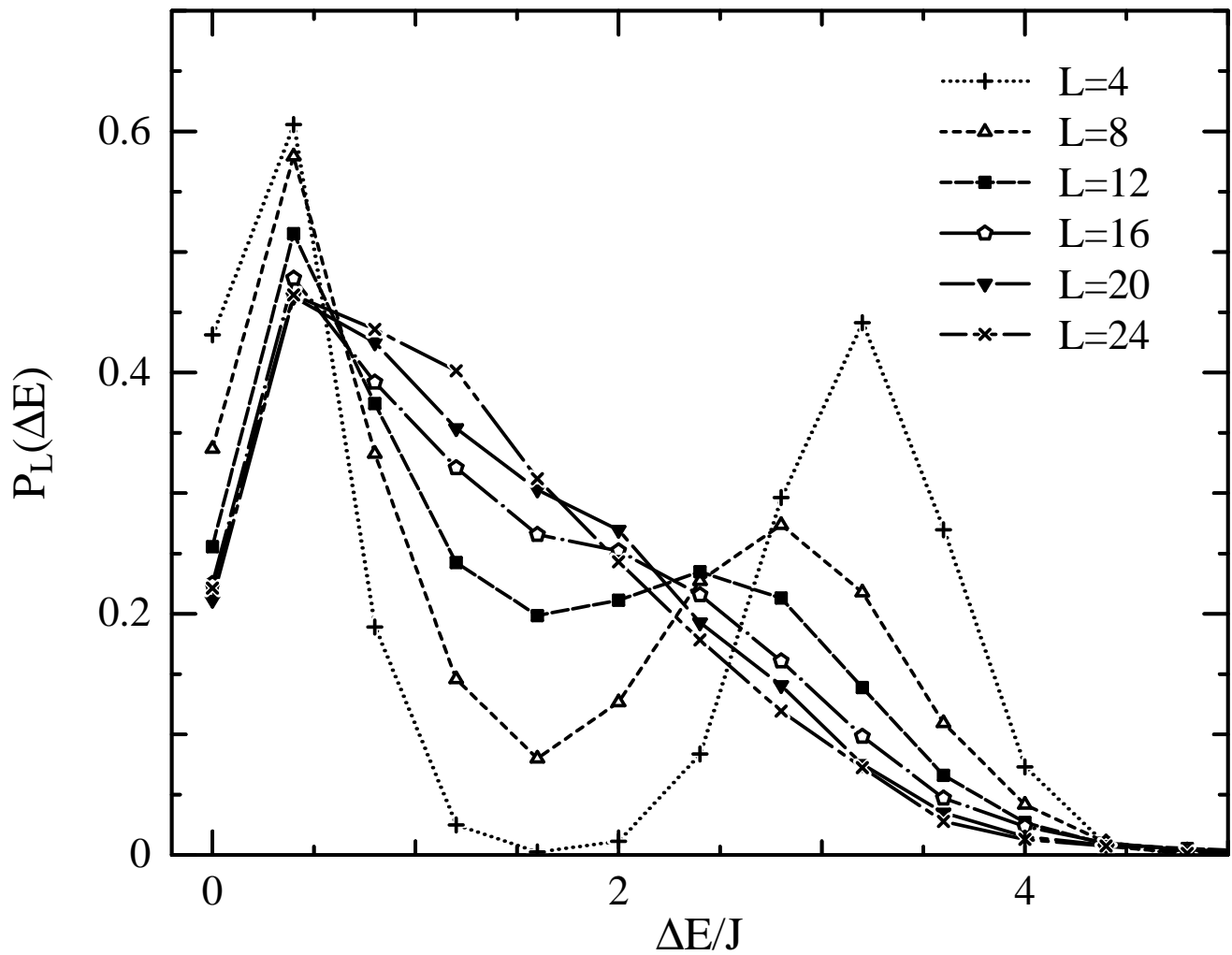
\begin{figure}
\caption{Distribution of $\Delta E$ of the $+J$ and $-0.8J$ model. 
         Lines are guide to the eye. }
\end{figure}

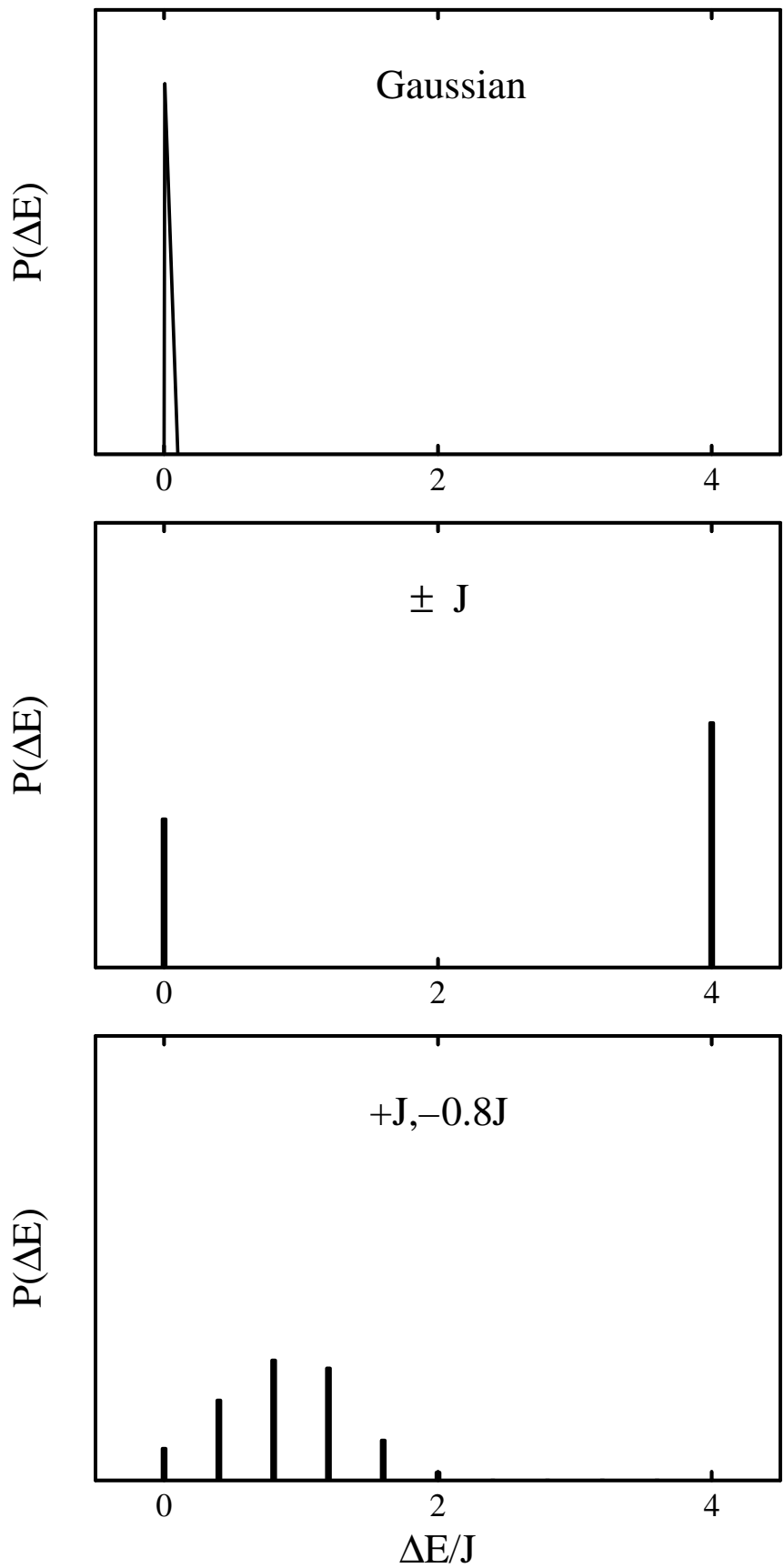
\begin{figure}
\caption{ Schematic pictures of distribution of $\Delta E$ for 
         $L \rightarrow \infty$.} 
\end{figure}

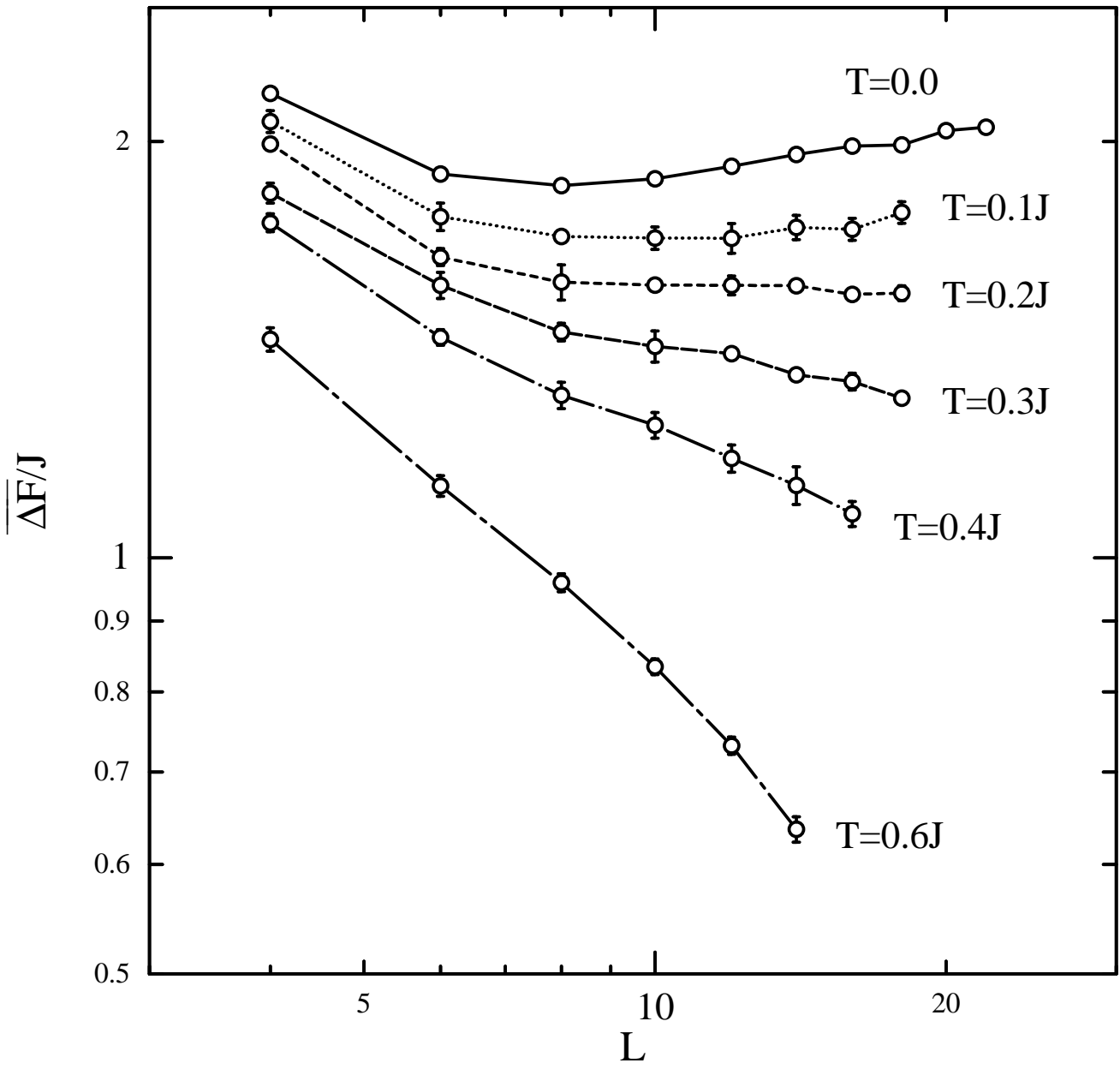
\begin{figure}
\caption{The interface free energies $\overline{\Delta F}$ of the $\pm J$ 
        model at different temperatures. Note that $\overline{\Delta F}$ 
        at $T = 0$ is $W(L)$. }
\end{figure}


\begin{references}
\bibitem{Comm2} We have also made two different evaluations of $\Delta E$. 
            One is to take an average of those for all \{$S_b$\} but a fixed 
            \{$S_a$\}. The other is to obtain the minimum value of it for 
            all sets of \{$S_a$\} and \{$S_b$\}. 
            The values of those evaluations are, of course, somewhat different 
            but their size dependences are similar to those presented here.
\bibitem{Comm3}  At finite temperatures, the sets of \{$S_a$\} and \{$S_b$\} 
            are determined in accordance with the Boltzmann's weight 
            \cite{SM2,CHB2}. For every sample, about 10 sets of them are used 
            to estimate $\Delta F$.
\end{references}
\end{document}